\documentclass[manuscript,screen]{acmart}
\AtBeginDocument{%
  \providecommand\BibTeX{{%
    \normalfont B\kern-0.5em{\scshape i\kern-0.25em b}\kern-0.8em\TeX}}}
\setcopyright{acmcopyright}
\copyrightyear{2023}
\acmYear{2023}
\acmDOI{}
\acmConference[Conference acronym 'XX]{The 1st International Workshop on Explainable AI for the Arts}{June 19, 2023}{}
\acmBooktitle{The 1st International Workshop on Explainable AI for the Arts (XAIxArts), ACM Creativity and Cognition (C\&C) 2023. Online, 3 pages} 
\acmPrice{15.00}
\acmISBN{978-1-4503-XXXX-X/18/06}

\begin{document}

\title{A Context-Sensitive Approach to XAI in Music Performance}

\author{Nicola Privato}
\email{nicola@lhi.is}
\author{Jack Armitage}
\email{jack@lhi.is}
\affiliation{%
  \institution{Intelligent Instruments Lab, Iceland University of the Arts}
  \streetaddress{Þverholt 11}
  \city{Reykjavik}
  \country{Iceland}
}

\renewcommand{\shortauthors}{Privato and Armitage, et al.}

\begin{abstract}
The rapidly evolving field of Explainable Artificial Intelligence (XAI) has generated significant interest in developing methods to make AI systems more transparent and understandable. However, the problem of explainability cannot be exhaustively solved in the abstract, as there is no single approach that can be universally applied to generate adequate explanations for any given AI system, and this is especially true in the arts. In this position paper, we propose an \textit{Explanatory Pragmatism} (EP) framework for XAI in music performance, emphasising the importance of context and audience in the development of explainability requirements. By tailoring explanations to specific audiences and continuously refining them based on feedback, EP offers a promising direction for enhancing the transparency and interpretability of AI systems in broad artistic applications and more specifically to music performance.
\end{abstract}




\received{1 May 2023}

\maketitle

\section{Introduction}

From a general perspective, the potential challenges for XAI include inherent lack of explainability of many systems, but also overwhelming complexity of an explanation, technical illiteracy of the audience, and a lack of a common frame of reference or relevant domain knowledge needed to grasp an explanation.
However, from this pragmatist account of AI explainability, addressing these challenges is not sufficient, particularly where musical performance is concerned.
Explainability cannot be exhaustively solved in the abstract, as no single approach to XAI can be applied off-the-shelf in order to generate adequate explanations for a given AI system. 
It becomes, therefore, necessary to consider the context in which the system operates.

In particular, explainability on this account should not be addressed per se, but in relation to a specific audience \cite{nyrup_robinson_2022}.
Responding to this issue, Nyrup and Robinson proposed their \textit{Explanatory Pragmatism} (EP) framework \cite{nyrup_robinson_2022} for the domain of medical AI, where they note that ``Without first specifying the relevant audience and purpose, there is no well-defined sense in which a given system is more or less explainable.''
To observe if and how the challenges of explainability apply to musical performative contexts, it is therefore necessary to establish the context in which explanation may or may not occur.
In the context of music , three main groups for AI explanation may be recognised: 

\begin{enumerate}
    \item An audience as a recipient of a musical outcome. 
    \item An ensemble of musicians and AI instruments.
    \item An individual musician interacting with an AI instrument.
\end{enumerate}

Here we consider the requirements for contextually-sensitive explanations for these three groups, from the perspective of music performance practice and culture, rather than science and engineering. 

\section{Target Groups for AI Explanation}

\subsection{XAI for Audiences of Musical Performances}

From the audience's perspective, what needs to be interpreted and/or explained is the musical and performative outcome in its entirety, rather than the predictions and actions of a single AI model. 
This is ontologically different from observing the explainability of an AI system as decoupled from the human agent, as the audience experiences the performative dialectics of the actors, the dance of agencies between the human and non-human participants whose interactions within the extended performative context characterise the musical outcome.

In this sense, explanatory pragmatism, when applied to the musical performance as interpreted by an audience, rather than to the algorithm may refer to collectives comprising human and machinic agents and to individual AI agents. 
These collectives are entities composed of people and computational agents, that work together to achieve a specific goal or set of goals. 
Cristianini defines such a \textit{Social Machine} as a ``machine where human participants and technical artefacts (e.g. a car, a piece of software, a robot) interact with one another to perform a task that would be hardly achievable by any single part'' \cite{cristianini2019social}.
Cristianini also defines Autonomous Social Machines (ASMs) that are driven by an intrinsic goal.
Music ensembles, particularly in contexts contemplating the use of agential or intelligent instruments, may be seen as a specific type of ASM having a common aesthetic goal.

In this context, relevant explanations might take into account the performative goals of the ensemble, address the extent to which the result is emergent or assigned, and in what measure it is consciously pursued by the human agents or depends on the AI algorithm interfacing the performers.

\subsection{XAI for Ensembles of Musicians and AI Instruments}

The integration of intelligent technologies in musical ensembles necessitates a re-evaluation of the dynamics between musicians and their instruments, as well as the explicit understanding of the evolving agency and expectations within the performance.

One of the most important concepts that students learn while playing in musical ensembles is that musicians are not supposed to play together at all times. Within each ensemble, it is possible to constitute smaller ensembles, such as dynamically articulating a jazz quartet into trio, duo, and solo moments during a performance. This approach notably increases the variability and enriches the experience for both the members and the audience.

Addressing AI explainability in this context requires providing an account of how the dynamics within an ensemble change throughout the performance. 
From the perspective of informative explanation, this could begin with typologies and topologies of performance ensembles that abstract them as networks of players, instruments, scores and audiences \cite{armitageAgentialScoresExploring2023}, and real-time analyses could be proposed to extend these displays temporally.
A cultural or aesthetic approach that complements this could involve interpreting the ensemble from the perspective of ritual \cite{morabitoRitualisticApproachSonic2022}.

\subsection{XAI for Musicians Playing AI Instruments}

John Coltrane describing his practice with Thelonious Monk and his tunes \cite{Coltrane}:

\begin{quote}
    ``I liked [‘Monk’s Mood’] so well I told him that I wanted to learn it, so he invited me around, and that’s when I started learning his tunes. I’d go by his apartment and I’d get him  out of bed maybe. And then he’d wake up and go to the piano and start playing. 
    [...]
    And he’d continue to play over and over and over and over-and I’d get this part. 
    [...]
    He’d rather a guy learn without reading because that way you feel it better. And so when I almost had the tune down, then he would leave me with it. He’d leave me to practice it alone.''
\end{quote}

In this anecdote, John Coltrane describes the process of learning Monk’s tunes. The explicit components are precise, pre-established harmonic and melodic structures. These are communicated, experienced and practiced through imitation rather than (in most cases) being read or explained. The goal is to turn the song's structure into a grid facilitating the communication between the performers. This grid needs to be felt rather than simply learnt.

In this context explaining is not sufficient, as musical information must enter the performer's embodied experience. Conceptual explanations in the form of melodies and chord progressions need to enter what Stiegler defines \textit{primary retention}, as the process by which our consciousness retains and processes the sensory data that we receive in the present moment \cite{Stiegler1998-STITAT-3}. 

Finally, in order to be disposable, the embodied musical information needs to be stabilised and processed through sustained practice, to the point that learning a 32 bars tune requires a whole day of practice, as reported by Coltrane. We argue that in this third explanatory context practice is the locus in which the embodied understanding of the musical form and of the instrument's agency should be framed. 

\section{Closing Remarks}

In this position paper we observed the EP framework within the context of music performance. We argued that in music performance three main contexts for explanations may be distinguished: The audience as a recipient of a musical outcome. The ensemble of musicians and AI instruments and the individual musician interacting with an AI instrument. We also reflected on what forms explanations might take in each of these.





\begin{acks}
This research is supported by the European Research Council (ERC) as part of the Intelligent Instruments project (INTENT), under the European Union's Horizon 2020 research and innovation programme (Grant agreement No. 101001848).
\end{acks}
\bibliographystyle{ACM-Reference-Format}
\bibliography{sample-base}


\begin{thebibliography}{6}


\ifx \showCODEN    \undefined \def \showCODEN     #1{\unskip}     \fi
\ifx \showDOI      \undefined \def \showDOI       #1{#1}\fi
\ifx \showISBNx    \undefined \def \showISBNx     #1{\unskip}     \fi
\ifx \showISBNxiii \undefined \def \showISBNxiii  #1{\unskip}     \fi
\ifx \showISSN     \undefined \def \showISSN      #1{\unskip}     \fi
\ifx \showLCCN     \undefined \def \showLCCN      #1{\unskip}     \fi
\ifx \shownote     \undefined \def \shownote      #1{#1}          \fi
\ifx \showarticletitle \undefined \def \showarticletitle #1{#1}   \fi
\ifx \showURL      \undefined \def \showURL       {\relax}        \fi
\providecommand\bibfield[2]{#2}
\providecommand\bibinfo[2]{#2}
\providecommand\natexlab[1]{#1}
\providecommand\showeprint[2][]{arXiv:#2}

\bibitem[Armitage and Magnusson(2023)]%
        {armitageAgentialScoresExploring2023}
\bibfield{author}{\bibinfo{person}{Jack Armitage} {and} \bibinfo{person}{Thor
  Magnusson}.} \bibinfo{year}{2023}\natexlab{}.
\newblock \showarticletitle{Agential {{Scores}}: {{Exploring Emergent}},
  {{Self-Organising}} and {{Entangled Music Notation}}}. In
  \bibinfo{booktitle}{\emph{Proceedings of the 8th {{International Conference}}
  on {{Technologies}} for {{Music Notation}} and {{Representation}}}}
  ({Northeastern University, Boston, Massachusetts, USA}, 2023).
\newblock


\bibitem[Cristianini and Scantamburlo(2019)]%
        {cristianini2019social}
\bibfield{author}{\bibinfo{person}{Nello Cristianini} {and}
  \bibinfo{person}{Teresa Scantamburlo}.} \bibinfo{year}{2019}\natexlab{}.
\newblock \bibinfo{title}{On Social Machines for Algorithmic Regulation}.
\newblock
\newblock
\showeprint[arxiv]{1904.13316}~[cs.CY]


\bibitem[Morabito et~al\mbox{.}(2022)]%
        {morabitoRitualisticApproachSonic2022}
\bibfield{author}{\bibinfo{person}{Robin Morabito}, \bibinfo{person}{Jack
  Armitage}, {and} \bibinfo{person}{Thor Magnusson}.}
  \bibinfo{year}{2022}\natexlab{}.
\newblock \showarticletitle{Ritualistic Approach to Sonic Interaction Design:
  {{A}} Poetic Framework for Participatory Sonification}.
  \bibinfo{publisher}{{Georgia Institute of Technology}}.
\newblock
\urldef\tempurl%
\url{https://doi.org/10.21785/icad2022.018}
\showDOI{\tempurl}


\bibitem[Nyrup and Robinson(2022)]%
        {nyrup_robinson_2022}
\bibfield{author}{\bibinfo{person}{Rune Nyrup} {and} \bibinfo{person}{Diana
  Robinson}.} \bibinfo{year}{2022}\natexlab{}.
\newblock \showarticletitle{Explanatory pragmatism: a context-sensitive
  framework for explainable medical AI}.
\newblock \bibinfo{journal}{\emph{Ethics and Information Technology}}
  \bibinfo{volume}{24}, \bibinfo{number}{1} (\bibinfo{date}{Feb}
  \bibinfo{year}{2022}).
\newblock
\urldef\tempurl%
\url{https://doi.org/10.1007/s10676-022-09632-3}
\showDOI{\tempurl}


\bibitem[Porter(1997)]%
        {Coltrane}
\bibfield{author}{\bibinfo{person}{Lewis Porter}.}
  \bibinfo{year}{1997}\natexlab{}.
\newblock \bibinfo{booktitle}{\emph{John Coltrane : his life and music}}.
\newblock \bibinfo{publisher}{University of Michigan Press}.
\newblock


\bibitem[Stiegler(1998)]%
        {Stiegler1998-STITAT-3}
\bibfield{author}{\bibinfo{person}{Bernard Stiegler}.}
  \bibinfo{year}{1998}\natexlab{}.
\newblock \bibinfo{booktitle}{\emph{Technics and Time, 2: Disorientation}}.
\newblock \bibinfo{publisher}{Stanford University Press}.
\newblock


\end{thebibliography}
\end{document}